\documentclass[pra,letterpaper,aps,10pt,twocolumn,floatfix]{revtex4-1}

\usepackage{amsmath, upgreek, amssymb, graphicx, paralist, gensymb}
\usepackage[colorlinks, linkcolor=blue, citecolor=blue, urlcolor=blue, breaklinks=true]{hyperref}

\newcommand{\aref}[1]{Appendix~\ref{#1}}
\newcommand{\eref}[1]{Eq.~(\ref{#1})}
\newcommand{\erefs}[1]{Eqs.~(\ref{#1})}

\newcommand{\Erefs}[1]{Equations~(\ref{#1})}
\newcommand{\fref}[1]{Fig.~\ref{#1}}
\newcommand{\sref}[1]{Sec.~\ref{#1}}
\newcommand{\Sref}[1]{Section~\ref{#1}}

\newcommand{\ie}{{i.e.}}

\newcommand{\eg}{{e.g.}}

\newcommand{\rmd}{\text{d}}

\newcommand{\abs}[1]{\lvert#1\rvert}

\renewcommand{\vec}[1]{\boldsymbol{#1}}
\newcommand{\mat}[1]{\boldsymbol{#1}}

\newcommand{\re}[1]{\,\text{Re}\!\left\{#1\right\}}

\makeatletter
\newlength \figurewidth
\makeatother

\begin{document}

\title{Selectable linear or quadratic coupling in an optomechanical system}

\date{\today}

\author{Andr\'e Xuereb}
\email[Corresponding author. E-mail address:\ ]{andre.xuereb@qub.ac.uk}
\affiliation{Centre for Theoretical Atomic, Molecular and Optical Physics, School of Mathematics and Physics, Queen's University Belfast, Belfast BT7\,1NN, United Kingdom}
\author{Mauro Paternostro}
\affiliation{Centre for Theoretical Atomic, Molecular and Optical Physics, School of Mathematics and Physics, Queen's University Belfast, Belfast BT7\,1NN, United Kingdom}

\begin{abstract}
There has been much interest recently in the analysis of optomechanical systems incorporating dielectric nano- or microspheres inside a cavity field. We analyse here the situation when one of the mirrors of the cavity itself is also allowed to move. We reveal that the interplay between the two oscillators yields a cross-coupling that results in, \eg, appreciable cooling and squeezing of the motion of the sphere, despite its nominal quadratic coupling. We also discuss a simple modification that would allow this cross-coupling to be removed at will, thereby yielding a purely quadratic coupling for the sphere.
\end{abstract}

\maketitle

Several early optomechanical systems~\cite{Arcizet2006,Gigan2006,Groblacher2009a} were straightforward realisations of ideas first proposed by Braginsky and Manukin~\cite{Braginsky1967,Braginsky1977}, and took the form of a Fabry--P\'erot cavity with one moving mirror. These systems are naturally modelled by the Hamiltonian~\cite{WilsonRae2007,Marquardt2007} (we will use units such that $\hbar=1$ throughout this paper) $\hat{H}={{\hat{p}}^2}/{2m}+m\omega_m^2\hat{x}^2/2+\hat{H}_\mathrm{lin}$, where we have introduced the term
\begin{equation}
\label{eq:Hlin}
\hat{H}_\mathrm{lin}=\bigl(\omega_\mathrm{c}-g_1\hat{x}\bigr)\hat{a}^\dagger\hat{a}
\end{equation}
with $\hat{a}$ the annihilation operator representing the cavity field (frequency $\omega_\mathrm{c}$), $\hat{x}$ [$\hat{p}$] the position [momentum] operator for the moving mirror (frequency $\omega_m$ and mass $m$), and $g_1$ the optomechanical coupling constant. Physically, $\hat{H}_{\textrm{lin}}$ admits a remarkably simple interpretation:\ the instantaneous position of the mirror changes the length of the cavity, and thus its resonance frequency.\par
The light--mirror interaction inherent in \eref{eq:Hlin} enables the passive cooling of the mechanical system, an effect that has been demonstrated in a vast array of experimental settings ~\cite{Arcizet2006,Gigan2006,Schliesser2006,Kleckner2006,Corbitt2007} and that has been progressively pushed to the point that optical cooling to the mechanical quantum ground state was recently achieved~\cite{Chan2011}. Moreover, it has long been recognised that the mechanical and optical fields can be prepared in an entangled steady-state~\cite{Vitali2007,Paternostro2007} by the means of \eref{eq:Hlin}. Bhattacharya and Meystre~\cite{Bhattacharya2007a} proposed a different paradigm, where the moving mirror was translucent and placed \emph{inside} the cavity, a configuration that was implemented in the seminal experiment reported in Refs.~\cite{Thompson2008,Sankey2010}. A remarkable feature of this model is that, under the right conditions and differently from the above, the light--mirror interaction may be described through an optomechanical Hamiltonian \emph{quadratic} in the position operator of the mechanical system and reading
\begin{equation}
\hat{H}_\mathrm{quad}=\bigl(\omega_\mathrm{c}+g_2\hat{x}^2\bigr)\hat{a}^\dagger\hat{a}\,,
\end{equation}
where $g_2$ is a coupling constant. This model has several interesting properties, including the possibility of performing QND measurements of the occupation number of the oscillator~\cite{Clerk2010}, but precludes cooling to leading order in the linearised regime achieved when the cavity field is populated by a very large number of photons~\cite{Bhattacharya2008} (see, however, Ref.~\cite{Nunnenkamp2010} for a treatment that goes beyond the linearised dynamics of the system).\par
\begin{figure}[b]
 \includegraphics[scale=0.5]{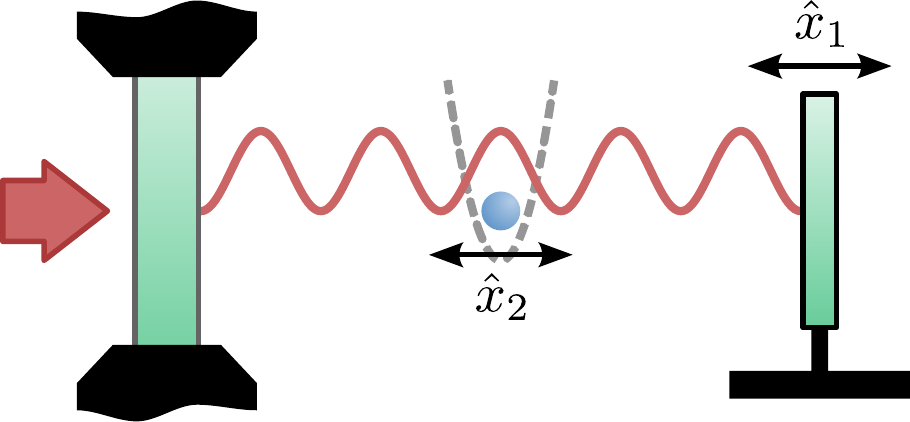}
 \caption{Schematic model of our system:\ a dielectric nano- or microparticle is coupled quadratically to a cavity field that is bounded by a moving mirror. The particle is held in place by an externally-generated harmonic potential, shown dotted, whose origin is left open in this work. The cavity field is pumped and decays only through the immobile mirror. Different pumping configurations are explored in \aref{app:Geometries}.}
 \label{fig:Model}
\end{figure}
In this paper we propose to combine these two scenarios, somewhat in the spirit of proposals that place a cloud of ultracold atoms inside an optomechanical setup~\cite{Paternostro2010,DeChiara2011,Rogers2012}. Let us consider the system shown in \fref{fig:Model}, which includes a small dielectric scatterer (herein referred to as the {\it sphere}) held in a harmonic potential inside a cavity~\cite{Chang2009b,RomeroIsart2010,RomeroIsart2011,RomeroIsart2011b,RomeroIsart2011c,Li2011,Monteiro2012}, one of whose end-mirrors is allowed to oscillate. As we shall show, this system allows us to exert a degree of control on the state of the motion of the sphere through its indirect interaction with the end-mirror, mediated by the cavity field. As a result of this, and by means of a judicious choice of operating parameters, we will show that passive cooling of the motion of the sphere, despite its apparent quadratic coupling to the cavity field, and squeezing of this same motion are both possible. Moreover, we expect that the interaction between the cavity field, sphere, and end-mirror gives rise to steady states with interesting genuinely tripartite entangled~\cite{Xuereb2012e} or nonlocal~\cite{Lee2012b} properties.\par
One might think that the Hamiltonian that describes this compound system would simply be given by the sum of the interaction terms in $\hat{H}_\mathrm{lin}$ through the position $\hat{x}_1$ of the end-mirror and $\hat{H}_\mathrm{quad}$ through the position $\hat{x}_2$ of the sphere. As natural as this might seem, this would not be entirely correct, given that the position of the end-mirror not only defines the resonance frequency of the cavity (as captured through the interaction in $\hat{H}_\mathrm{lin}$), but also determines the relative position of the cavity field (anti)nodes and the sphere. This means that, for the configuration shown in~\fref{fig:Model}, \ie, for a cavity pumped from the leftmost (fixed) mirror, the light field couples quadratically to the relative position operator $\hat{x}_1-\hat{x}_2$. We can thus write the Hamiltonian
\begin{equation}
\label{eq:Hlp}
\hat{H}_\mathrm{lp}=\Bigl[\omega_\mathrm{c}-g_1\hat{x}_1+g_2\bigr(\hat{x}_1-\hat{x}_2\bigr)^2\Bigr]\hat{a}^\dagger\hat{a}\,,
\end{equation}
where the subscript `lp' reminds us that the system is pumped from the left. In \aref{app:Geometries}, we examine the other possibilities for the geometry of the pump field.\par
The remainder of this paper is structured as follows. \Sref{sec:EoM} will set the stage for exploring the dynamics of our system, deriving in particular the Hamiltonian and the linearised equations of motion. \Sref{sec:Cooling} subsequently makes use of numerical simulations to discuss the resulting dynamics. As an application of our scheme, we show that the cross-coupling between the two oscillators, cf.~\eref{eq:Hlp}, gives rise to efficient cooling of the mechanical motion of the sphere. In \sref{sec:Squeezing} we discuss squeezing of the motion of the sphere and demonstrate that this motion can be squeezed appreciably if the coupling constant $g_2$ is large enough. Finally, we conclude with an outlook of possible future directions.

\section{Dynamics in the linearised regime}\label{sec:EoM}
Let us start by considering the model in~\fref{fig:Model}. Classically, the intensity of the field within the cavity at a distance $z$ from the right mirror is given by $I(z)=I_0\sin^2(kz)$, 
where the $I_0$ depends on the value of the wavenumber $k$ and the amplitude of the input field. To treat our oscillators quantum-mechanically we perform the replacement
\begin{equation}
z\to\hat{x}_1-\hat{x}_2+d,
\end{equation}
where $d$ is the distance between the equilibrium positions of the two oscillators. Let us choose $d$ such that $kd=n\lambda/2~(n\in\mathbb{Z})$, with $\lambda=2\pi/k$ the wavelength of the input light. With this choice, the sphere oscillates about a node or an antinode of the cavity field. The interaction Hamiltonian involving the sphere thus takes the form
\begin{equation}
\label{eq:Hint}
\hat{H}\propto\sin^2\bigl[k\bigl(\hat{x}_1-\hat{x}_2\bigr)\bigr]\hat{a}^\dagger\hat{a}.
\end{equation}
We now expand \eref{eq:Hint} as a series, stopping at the lowest order, introduce dimensionless mechanical operators [obtained by dividing out the extent of the zero-point fluctuations $x_{0,j}=\sqrt{{\hbar}/(m_j\omega_j)}$ of each oscillator, where $m_j$ is the mass and $\omega_j$ the oscillation frequency of the mirror ($j=1$) or the sphere ($j=2$)], and add back the linear optomechanical coupling between $\hat{a}$ and $\hat{x}_1$. This gives us the total interaction Hamiltonian
\begin{equation}
\hat{H}_\mathrm{int}=\Bigl[-g_1\hat{x}_1+g_2\bigl(\chi\hat{x}_1-\hat{x}_2\bigr)^2\Bigr]\hat{a}^\dagger\hat{a},
\end{equation}
where, as before, $g_1$ and $g_2$ are optomechanical coupling constants and $\chi=x_{0,1}/x_{0,2}$. The signs in front of the coupling constants are chosen to aid interpretation:\ the system shown in \fref{fig:Model} has positive $g_1$, and $g_2>0$ implies that the sphere oscillates about an antinode, \ie, a point of stable equilibrium. A full description of the system requires also the free Hamiltonian, which we write in a frame rotating at the frequency of the driving laser $\omega_\mathrm{L}$,
\begin{equation}
\hat{H}_\mathrm{free}=-\Delta\hat{a}^\dagger\hat{a}+\frac12\sum_{j=1}^2{\omega_j}\bigl(\hat{x}_j^2+\hat{p}_j^2\bigr)\,,
\end{equation}
as well as cavity-pump ($\hat{H}_\mathrm{pump}$) and dissipation Hamiltonians ($\hat{H}_\mathrm{diss}$), which will be left unspecified. Here, $\Delta=\omega_\mathrm{L}-\omega_\mathrm{c}$ is the cavity--pump detuning. The full Hamiltonian is thus $\hat{H}=\hat{H}_{\textrm{int}}+\hat{H}_{\textrm{free}}+\hat{H}_{\textrm{diss}}+\hat{H}_{\textrm{pump}}$. The corresponding equation of motion for the cavity field operator is 
\begin{equation}
\dot{\hat{a}}=\bigl(i\Delta-\kappa_\mathrm{c}\bigr)\hat{a}+i\Bigl[g_1\hat{x}_1-g_2\bigl(\chi\hat{x}_1-\hat{x}_2\bigr)^2\Bigr]\hat{a}-\sqrt{2\kappa_\mathrm{c}}\hat{a}_\mathrm{in}\,,
\end{equation}
where $\kappa_\mathrm{c}$ is the cavity amplitude decay rate, and $\hat{a}_\mathrm{in}$ is the input field. As for the sphere and mechanical mode, the dynamics is encompassed by
\begin{align}
\label{eq:Dynamics}
\begin{split}
\dot{\hat{x}}_j&=\omega_j\hat{p}_j,\\
\dot{\hat{p}}_1&=-\omega_1\hat{x}_1+(g_1-\chi\hat{\cal{X}}
)\hat{a}^\dagger\hat{a}-2\gamma_1\hat{p}_1-\sqrt{2\gamma_1}\hat{\xi}_1,\\
\dot{\hat{p}}_2&=-\omega_2\hat{x}_2+\hat{\cal{X}}\hat{a}^\dagger\hat{a}-2\gamma_2\hat{p}_2-\sqrt{2\gamma_2}\hat{\xi}_2.
\end{split}
\end{align}
where $\hat{\cal{X}}=2g_2(\chi\hat{x}_1-\hat{x}_2)$, $\gamma_j$ are the mechanical decay rates, and $\hat{\xi}_j$ is the self-adjoint zero-mean operator describing the Brownian noise affecting oscillator $j=1,2$ and characterised by the correlation functions~\cite{Giovannetti2001}
\begin{equation}
\langle\hat{\xi}_j(t)\hat{\xi}_l(t^\prime)\rangle=(2n_j+1)\delta_{jl}\delta(t-t^\prime)\,,
\end{equation}
where $n_j=\left[e^{\hbar\omega_j/(k_\mathrm{B}T_j)}-1\right]^{-1}$ is the equilibrium phononic population of oscillator $j$ at temperature $T_j$ ($k_\mathrm{B}$ is the Boltzmann constant).

\subsection{Classical steady-state solution}
\Erefs{eq:Dynamics} are tackled in two steps. First we replace each operator $\hat{o}$ by its quantum-mechanical mean value $\bar{o}$. The resulting equations of motion are identical to the classical ones. Their solution is then used in a second step, as described in~\sref{sec:EoM:Linear}, to analyse the linear evolution of the system. We remark here that $\bar{\xi}_j=0~~(j=1,2)$, and that $\bar{a}_\mathrm{in}$ relates to the input power as $P_\mathrm{in}=\hbar\omega_\mathrm{L}\abs{\bar{a}_\mathrm{in}}^2$. When the dynamics is stable, the long-time classical solutions are
\begin{equation}
\label{eq:Classical}
\bar{a}=\frac{\sqrt{2\kappa_\mathrm{c}}}{i\tilde{\Delta}-\kappa_\mathrm{c}}\bar{a}_\mathrm{in},\ \bar{x}_2=\bar{x}_1\frac{2g_2\abs{\bar{a}}^2}{\Omega_2}=\frac{2g_1g_2\chi\abs{\bar{a}}^4}{\Omega_1\Omega_2},
\end{equation}
where we have replaced the detuning $\Delta$ with $\tilde{\Delta}=\Delta+g_1\bar{x}_1-g_2\bigl(\chi\bar{x}_1-\bar{x}_2\bigr)^2$ and introduced the quantities
\begin{align}
\begin{split}
\Omega_1&=\omega_1+2g_2\chi^2\abs{\bar{a}}^2-\frac{4g_2^2\chi^2\abs{\bar{a}}^4}{\omega_2+2g_2\abs{\bar{a}}^2},\\
\Omega_2&=\omega_2+2g_2\abs{\bar{a}}^2.
\end{split}
\end{align}
\Erefs{eq:Classical} diverge in a notable way from the analogous ones reported in Ref.~\cite{Bhattacharya2008}. The latter case corresponds to ours upon going to the infinite-mass limit for the end-mirror ($\chi=g_1=0$). From \erefs{eq:Classical} it then follows that $\bar{x}_2=\bar{x}_1=0$. An immediate consequence of this is that in the linearised equations of motion, as obtained in the next Section, the motion of the mirror is decoupled from the rest of the system [cf.~\eref{eq:DriftMatrix}]. Under these conditions, therefore, the effective mechanical decay rate of the sphere cannot be altered by the dynamics in the linear approximation, \ie, the motion of the sphere cannot be cooled by a quadratic interaction with the cavity field. Outside the infinite-mass limit, however, the two oscillators interact to avoid this pitfall. As we shall see later on, one can choose the parameters such that cooling of the sphere to a temperature determined mostly by the properties of the end-mirror is indeed possible, thus demonstrating the control mechanism at the centre of our work.

\subsection{Linearised equations of motion}\label{sec:EoM:Linear}
Having at hand the explicit form of the operators' mean values, we now consider the zero-mean quantum fluctuations $\delta\hat{o}=\hat{o}-\bar{o}$. When $\abs{\bar{o}}^2\gg1$ one can truncate the equations of motion to first order in the fluctuations, yielding the set of linear equations of motion
\begin{equation}
\label{eq:EoMLin}
\frac{\rmd}{\rmd t}\hat{\vec{R}}=\mat{A}\,\hat{\vec{R}}+\hat{\vec{R}}_\mathrm{in}\,.
\end{equation}
Here, we have defined the vectors of quadrature fluctuations $\hat{\vec{R}}=\bigl(\delta\hat{x},\delta\hat{p},\delta\hat{x}_1,\delta\hat{p}_1,\delta\hat{x}_2,\delta\hat{p}_2\bigr)^\mathrm{T}$, where $\delta\hat{x}=({\delta\hat{a}+\delta\hat{a}^\dagger})/{\sqrt{2}}$ and $\delta\hat{p}=i({\delta\hat{a}^\dag-\delta\hat{a}})/{\sqrt{2}}$,
as well as $\hat{\vec{R}}_\mathrm{in}=-\bigl(\sqrt{2\kappa_\mathrm{c}}\delta\hat{x}_\mathrm{in},\sqrt{2\kappa_\mathrm{c}}\delta\hat{p}_\mathrm{in},0,\sqrt{2\gamma_1}\hat{\xi}_1,0,\sqrt{2\gamma_1}\hat{\xi}_2\bigr)^\mathrm{T}$ with $\delta\hat{x}_\mathrm{in}$ and $\delta\hat{p}_\mathrm{in}$ that are defined analogously to $\delta\hat{x}$ and $\delta\hat{p}$. The input noise field obeys $\langle\delta\hat{a}_\mathrm{in}(t)\delta\hat{a}_\mathrm{in}(t^\prime)\rangle=\langle\delta\hat{a}_\mathrm{in}(t)^\dagger\delta\hat{a}_\mathrm{in}(t^\prime)\rangle=0$ and $\langle\delta\hat{a}_\mathrm{in}(t)\delta\hat{a}_\mathrm{in}^\dagger(t^\prime)\rangle=\delta(t-t^\prime)$. Moreover, we have introduced the drift matrix $\mat{A}$, whose (lengthy) expression is given in \aref{app:Drift}. Stability of the dynamics encompassed by \eref{eq:EoMLin} is easily studied by using the well-known Routh--Hurwitz criterion~\cite[\S 2.3.2]{Horn2008}, which (in its most straightforward formulation) guarantees the existence of a stable solution if all of the eigenvalues of $\mat{A}$ have negative real part. All of the following analysis is conducted in this stable regime.\par
The linearised dynamics for the operator increments can be derived from a quadratic effective Hamiltonian. For a fixed set of parameters, the dynamics therefore leads any initial Gaussian state to a unique final Gaussian steady state. This steady state has zero first moment, since we operate in a shifted picture involving only zero-mean operator increments, and the matrix of second moments $\mat{V}=[\langle\vec{R}(t)\otimes\vec{R}(t^\prime)\rangle+\langle\vec{R}(t)\otimes\vec{R}(t^\prime)\rangle^\mathrm{T}]/2$ satisfies the Lyapunov equation $\mat{A}\,\mat{V}+\mat{V}\,\mat{A}^\mathrm{T}=-\boldsymbol{D}$, where
\begin{equation*}
\boldsymbol{D}\delta(t-t^\prime)=\tfrac{1}{2}\bigl[\langle\vec{R}_\mathrm{in}(t)\otimes\vec{R}_\mathrm{in}(t^\prime)\rangle+\langle\vec{R}_\mathrm{in}(t)\otimes\vec{R}_\mathrm{in}(t^\prime)\rangle^\mathrm{T}\bigr]\,.
\end{equation*}
We remark that the Lyapunov equation admits a compact solution in in terms of the eigensystem of $\boldsymbol{A}$~\cite[\S 5.0.4]{Horn2008}. Having at hand the covariance matrix $\mat{V}$, we can readily analyse the (Gaussian) steady state of the system, including bipartite entanglement between any two components of the system, genuine tripartite entanglement, or occupation numbers for the two harmonic oscillators.

\begin{figure}[t]
 \includegraphics[width=\figurewidth]{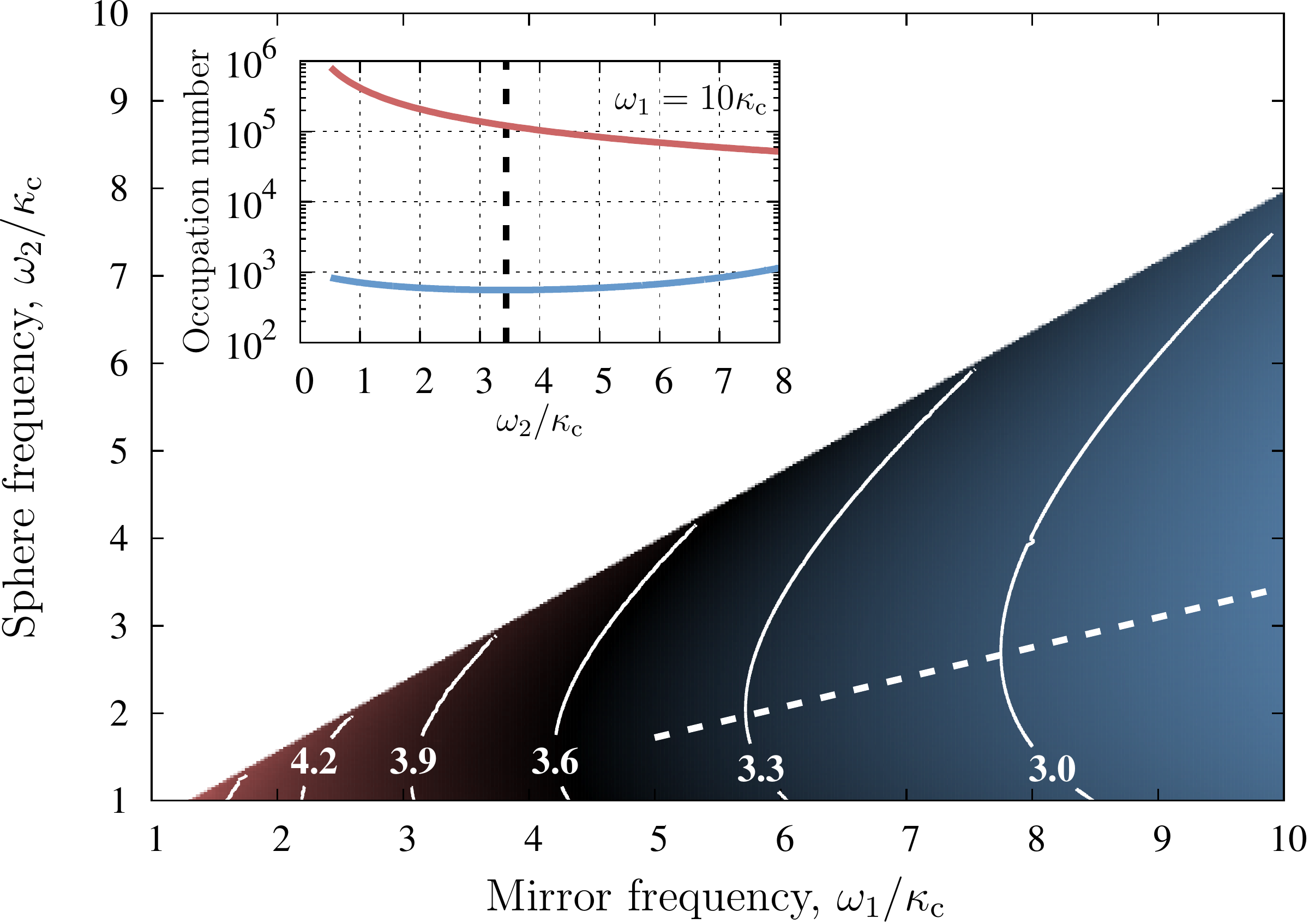}
 \caption{Occupation number for the motion of the sphere, presented on a logarithmic scale. Note the `resonant' condition, indicated by the dashed line, corresponding to the sphere frequency that minimises this occupation number at each particular $\omega_1$. Cooling of the motion of the sphere by almost three orders of magnitude is seen, despite the nominal quadratic coupling. The plot here has been optimised over $\tilde{\Delta}$ and $P_\mathrm{in}$. The mirror and sphere are in contact with separate baths at temperatures of $50$\,mK and $1$\,K, respectively. The rest of the parameters used for generating this plot are in \sref{sec:Cooling}. \emph{Inset:}~Cut of the main figure along the line $\omega_1=10\kappa_\mathrm{c}$ (blue, lower, curve). The red, upper, curve shows the occupation number in the absence of any cooling. The dashed line indicates the value of $\omega_2$ for which the lowest occupation number is reached.}
 \label{fig:Occupation}
\end{figure}
\begin{figure}[t]
 \rlap{(a)}{\vspace{-0.5em}
\includegraphics[width=\figurewidth]{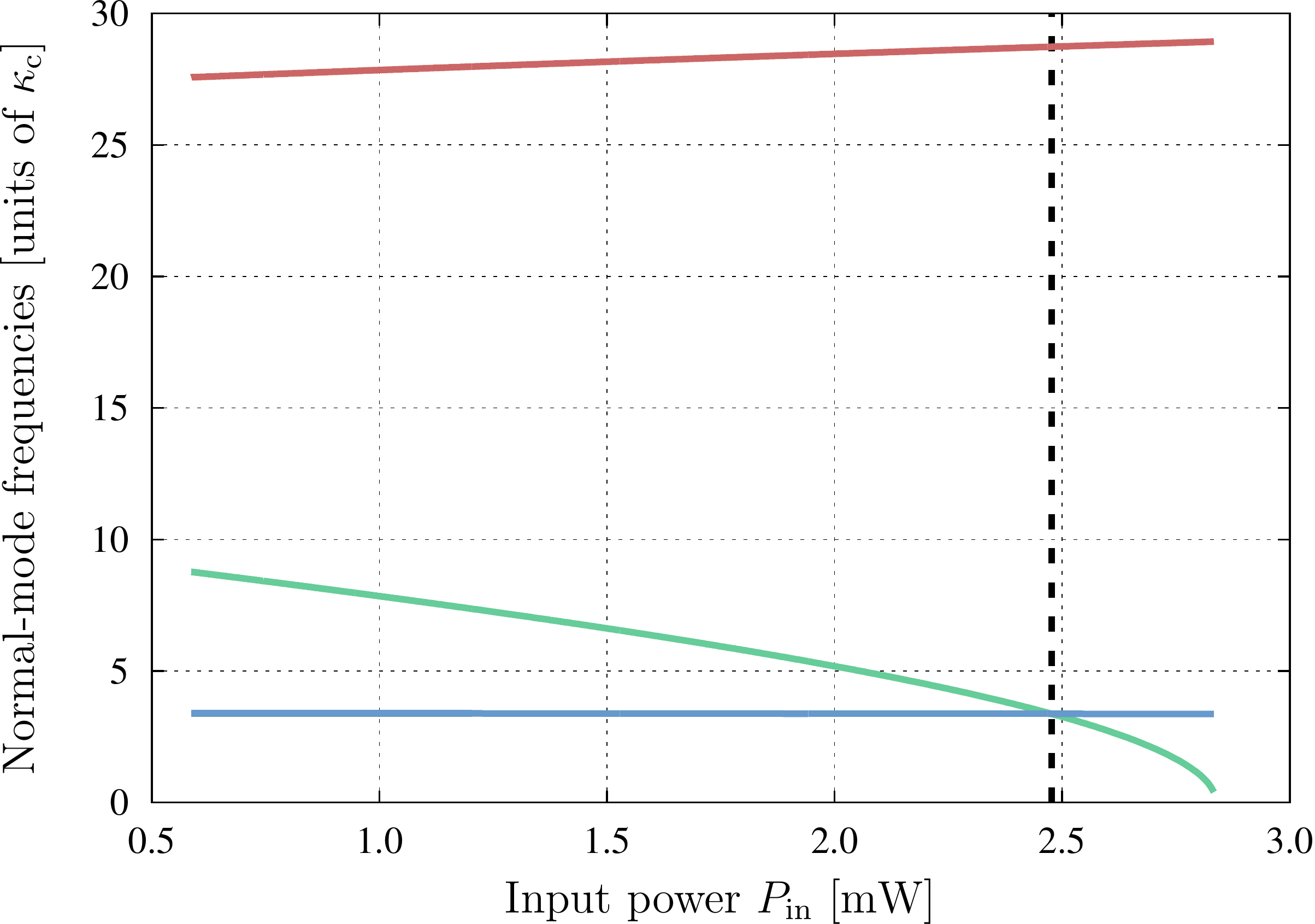}
 }\\
 \rlap{(b)}{\vspace{-0.5em}
 \includegraphics[width=\figurewidth]{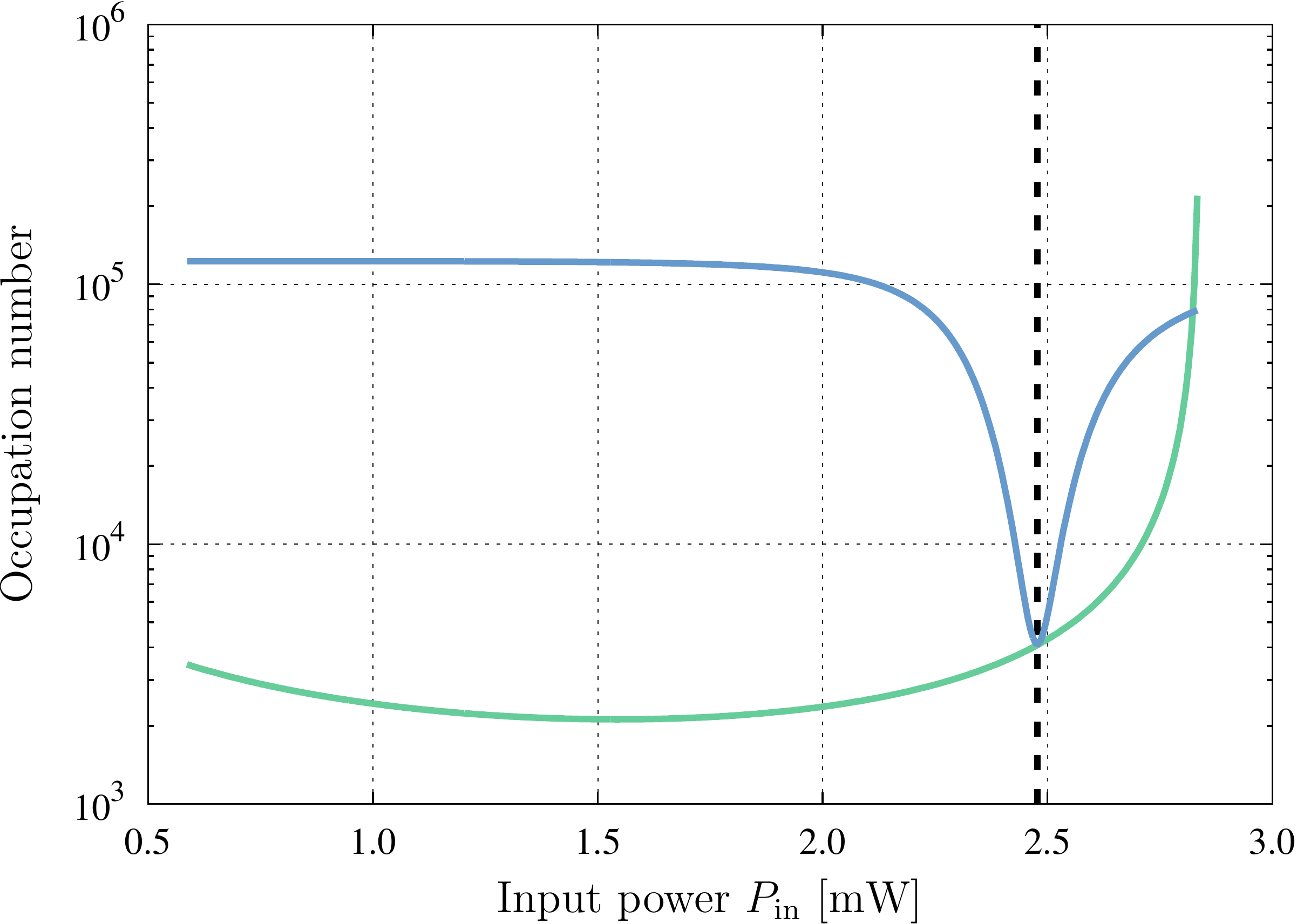}
 }
 \caption{(a)~Normal-mode frequencies, extracted from the eigenvalues of $\boldsymbol{A}$. For low powers, the red (top) curve represents the cavity mode, the green (middle, at low powers) curve the mirror, and the blue (bottom) curve the sphere. (b)~Occupation numbers for the sphere (blue; upper curve at low powers) and mirror (green; lower). Note the narrow range of powers for which the systems hybridise. At this point, the two frequencies and occupation numbers are equal. We have chosen $\omega_1=10\kappa_\mathrm{c}$ and the optimal values for $\omega_2$ and $\tilde{\Delta}$. The mirror and sphere are both in contact with baths at a temperature of $1$\,K. The dashed line denotes the value of $P_\mathrm{in}$ for which the occupation number is minimised. ($\omega_1=10\kappa_\mathrm{c}$, $\omega_2=3.4\kappa_\mathrm{c}$, $\tilde{\Delta}=-27.2\kappa_\mathrm{c}$, $g_1=1.0\times10^{-3}\kappa_\mathrm{c}$, $g_2=-2.4\times10^{-10}\kappa_\mathrm{c}$, $\chi=3.7\times10^{-3}$; other parameters as in the body of the paper.)}
 \label{fig:FrequenciesOccupations}
\end{figure}
\section{Cooling of the sphere}\label{sec:Cooling}
In this Section, we shall demonstrate the claimed cooling mechanism for the trapped sphere, and the control we can operate on it through the means of the mechanical mirror. Before specifying the values of the various parameters that enter our dynamical equations, let us give explicit formulae for $g_1$ and $g_2$. The basic theory of cavity optomechanics identifies $g_1$ with the first-order change in the cavity resonance frequency when the end-mirror moves by a distance $x_{0,1}$~\cite{Kippenberg2007}. For a cavity of length $L$ we can thus write~\cite{Xuereb2012d}
\begin{equation}
g_1=\frac{\omega_\mathrm{c}}{L}x_{0,1}=\frac{\omega_\mathrm{c}}{L}\sqrt{\frac{\hbar}{m_1\omega_1}}\,.
\end{equation}
For $g_2$, in the case of spheres that are not too large, we use the expression in Ref.~\cite{RomeroIsart2011c}
\begin{equation}
g_2=\pm\frac{3V}{4V_\mathrm{c}}\re{\frac{\epsilon_\mathrm{r}-1}{\epsilon_\mathrm{r}+2}}x_{0,2}^2k^3c\,,
\end{equation}
where $V$ is the volume of the sphere, $V_\mathrm{c}=\tfrac{\pi}{4}w^2L$ is the mode volume for a cavity with waist $w$, $k=2\pi/\lambda$ is the wavenumber of the driving field, and $\lambda$ its wavelength. The sign of this expression is chosen depending on whether the sphere is at a node~($-$) or an antinode~($+$) of the cavity field. In the case of a lossless dielectric of real refractive index $n$, $\re{\epsilon_\mathrm{r}}=n^2$, and we can write this expression as
\begin{equation}
g_2=\pm12\pi\frac{n^2-1}{n^2+2}\frac{\omega_\mathrm{c}}{L}\frac{\hbar}{\rho(\lambda w)^2\omega_2}\,,
\end{equation}
where $\rho$ is the mass density of the material composing the sphere. Note, in particular, that $g_2$ is independent of the radius $r$ of the sphere. For silica $\rho=2650$\,kg\,m$^{-3}$ and, in the near infrared, $n\approx1.5$. We shall use $\lambda=1064$\,nm, $L=0.5$\,cm, $\kappa_\mathrm{c}=2\pi\times50$\,kHz, $r=0.5$\,$\upmu$m, $w=40$\,$\upmu$m~(which can all be found in Ref.~\cite{Monteiro2012}), and $m_1=40$\,ng~\cite{Groblacher2009b}. For the mechanical damping constants, we take the values $\gamma_1=2\pi\times140$\,Hz~\cite{Groblacher2009a} and $\gamma_2=2\pi\times0.5$\,mHz. At the two nominal frequencies $\omega_1=1$\,MHz and $\omega_2=2\pi\times200$\,kHz, we find $g_1\approx2\pi\times36$\,Hz, $g_2\approx-2\pi\times10$\,$\upmu$Hz at a node of the cavity field, and $\chi\approx2.9\times10^{-2}$. In what follows we will not constrain ourselves to these particular values of $\omega_j$, however, and will scale $g_j$ and $\chi$ appropriately. In passing, we note that $g_1$ and $g_2$ can be made larger by decreasing the length of the cavity or lowering the mechanical oscillation frequencies. Moreover, $g_1$ can be increased by lowering the effective mass of the moving mirror, while $g_2$ by using a sphere with a larger refractive index or a cavity with a smaller waist. Finally, the value of $\chi$ can be made larger by increasing the ratios $\omega_2/\omega_1$ and $m_2/m_1$.\par
We have already mentioned how $\chi=0$ gives rise to the steady-state solution $\bar{x}_2=0$~\cite{Bhattacharya2008}. In the linearised regime, the motion of the sphere is thus decoupled from that of the dynamics of the cavity--mirror system. This can easily be seen by substituting $\chi=0$ and $\bar{x}_2=0$ into the drift matrix $\boldsymbol{A}$. A straightforward method to explore the implications of the coupling induced by our interaction Hamiltonian is thus to look at cooling of the motion of the sphere. A sample of such data is shown in \fref{fig:Occupation}. The cases (i)~$\omega_2\gtrsim\omega_1$ or (ii)~$\omega_2\lesssim\kappa_\mathrm{c}$ have been excluded from the plot for reasons that will be clarified later on. We see that, for each $\omega_1$, there exists a band of frequencies $\omega_2$ for which the steady-state occupation for the spherical motion is significantly below the starting value. Moreover, for a particular $\alpha_\mathrm{opt}=\omega_1/\omega_2$ that depends most strongly on the properties of the mirror (in particular, the constant parameter $g_1\sqrt{\omega_1}$), the motion of the mirror and sphere equilibrate to the same occupation number. The data displayed in \fref{fig:Occupation} show the reduction of the occupation number of the sphere by almost three orders of magnitude. However, in our numerical exploration we have seen evidence that this is not an upper limit to the efficiency of the process. Indeed, decreasing the temperature of the thermal bath to which the mirror is coupled will yield a correspondingly lower occupation number for the sphere, as a result of dynamical equilibration of the sphere through the mechanical mirror.\par
The basic mechanism through which this cooling process occurs is a hybridisation of the motion of the two oscillators. For moderately large values of $\omega_1/\omega_2$, increasing the input power from zero causes the frequencies of the normal modes of the system, given by the imaginary parts of the eigenvalues of $\boldsymbol{A}$, to shift. Thus, for instance, the mode that at $P_\mathrm{in}=0$ describes the mirror motion shifts to a smaller frequency $\tilde{\omega}_1$~\cite{Genes2008}. At the point where $\tilde{\omega}_1\approx\omega_2$, the two modes hybridise and cooling ensues. For $\omega_1/\omega_2<1$, this hybridisation is no longer possible. However, if $\omega_1/\omega_2$ is too large, the system reaches instability before the criterion $\tilde{\omega}_1\approx\omega_2$ can be satisfied. Conversely, if $\omega_1/\omega_2$ is not large enough, the power at which $\tilde{\omega}_1\approx\omega_2$ is satisfied is too low for efficient cooling. Such hybridisation can be seen in \fref{fig:FrequenciesOccupations}, where we plot the case $\omega_1=10\kappa_\mathrm{c}$, at the optimal values for $\omega_2$ and $\tilde{\Delta}$.\par
Another unique feature of our system is that, after cooling the sphere, we can switch to coupling light into the cavity from the moving mirror. As explained in \aref{app:Geometries}, this means that the Hamiltonian that couples the motion of the sphere to the optical field is purely quadratic. This renders possible, for example, quantum non-demolition (QND) measurements of the energy of the harmonic oscillator embodied by the sphere. This protocol requires that the cavity can be pumped from either end-mirror, but is deterministic and otherwise less demanding than, \eg, measurement-based schemes for achieving the same~\cite{Vanner2011}.

\begin{figure}[t]
 \rlap{(a)}{\vspace{-0.5em}
\includegraphics[width=\figurewidth]{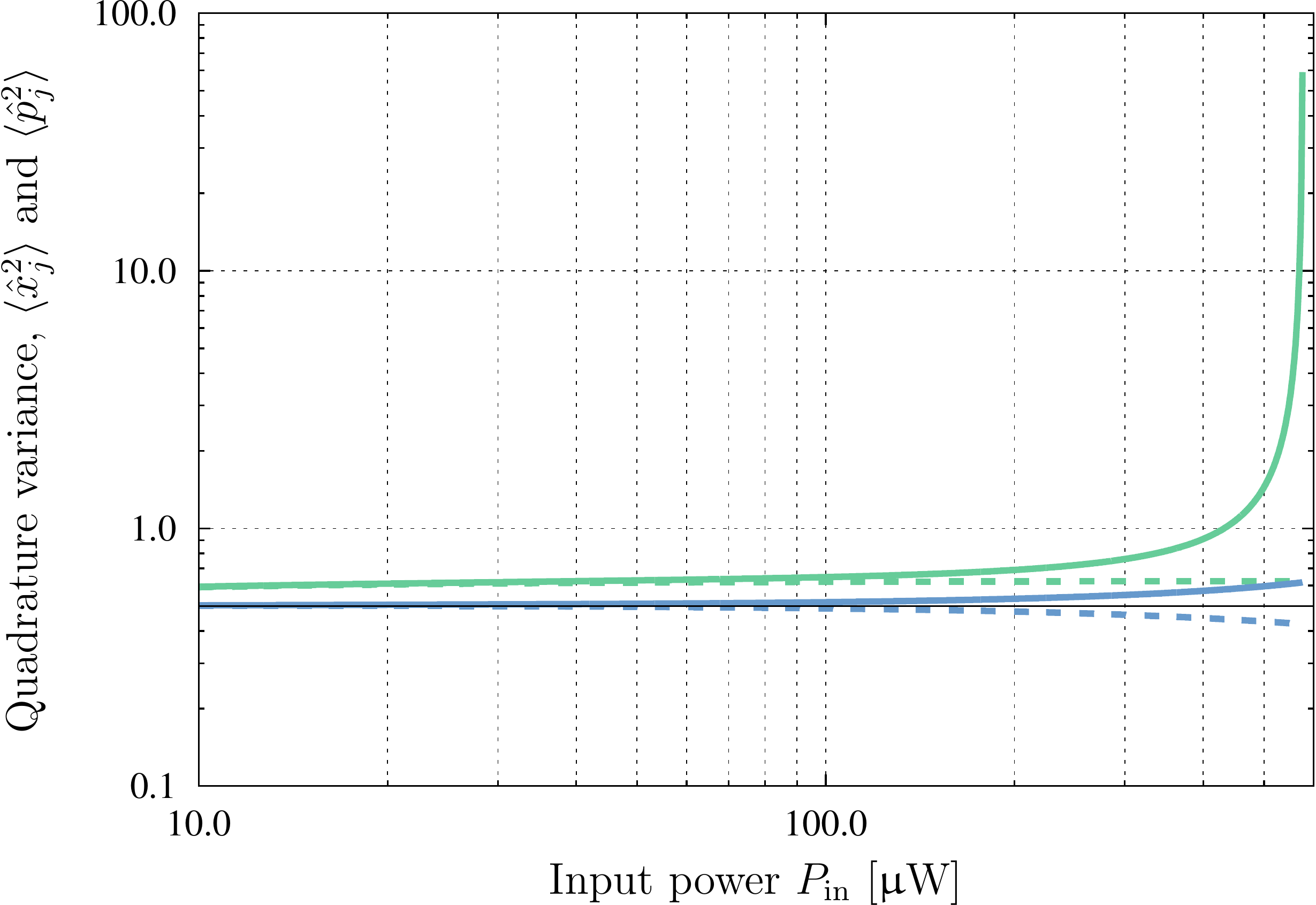}
 }\\
 \rlap{(b)}{\vspace{-0.5em}
 \includegraphics[width=\figurewidth]{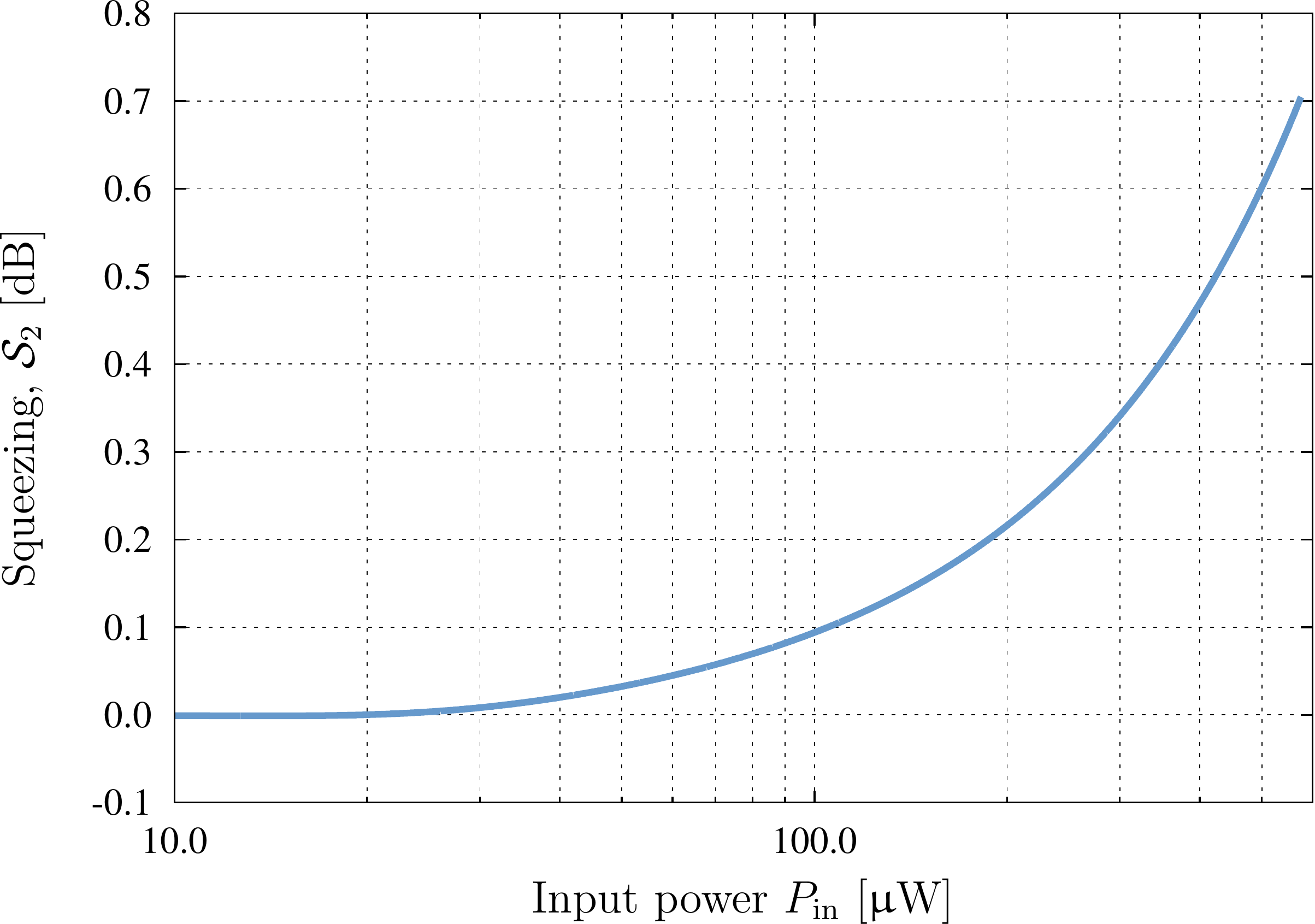}
 }
 \caption{(a)~Quadrature variances, $\langle\hat{x}_j^2\rangle$ (solid curves) and $\langle\hat{p}_j^2\rangle$ (dashed curves); the green, upper, curves represent the mirror ($j=1$), the blue, lower, curves the sphere ($j=2$). The solid black horizontal line represents the $\tfrac{1}{2}$-quantum ground-state vacuum variance. Note that only the $p$-quadrature of the sphere is squeezed below the vacuum level. The baths for both oscillators were held at zero temperature, and $g_2$ was a factor of $100$ larger than the parameters in the body of the manuscript. (b)~Corresponding squeezing for the sphere. ($\omega_1=20\kappa_\mathrm{c}$, $\omega_2=10\kappa_\mathrm{c}$, $\tilde{\Delta}=-10\kappa_\mathrm{c}$, $g_1=7.2\times10^{-4}\kappa_\mathrm{c}$, $g_2=-8.0\times10^{-9}\kappa_\mathrm{c}$, $\chi=4.5\times10^{-3}$; other parameters as in the body of the manuscript.)}
 \label{fig:Squeezing}
\end{figure}
\section{Squeezing of mechanical motion}\label{sec:Squeezing}
The coupling of mechanical elements to a light field, both linearly~\cite{Mancini1994} and quadratically~\cite{Nunnenkamp2010}, can give rise to squeezing of the mechanical motion, where the quantum nature of the optomechanical interaction acts to suppress the noise in one of the quadratures of the mechanical motion. To investigate the occurrence of mechanical squeezing in our setup we set the temperature of the baths coupled to both oscillators to zero. For the parameters given above, there is no observable squeezing in the motion of the sphere, and to induce such effects we have found it necessary to use a larger value for $g_2$, \eg, two orders of magnitude larger in the case of the data presented in \fref{fig:Squeezing}.\par
To quantify the amount of squeezing in the steady state of the linearised dynamics, we used the figure of merit
\begin{equation}
\mathcal{S}_j=\frac{1}{2\min\bigl\{\langle\hat{x}_j^2\rangle,\langle\hat{p}_j^2\rangle\bigr\}}\,,
\end{equation}
as plotted in \fref{fig:Squeezing}(b) for $j=2$. In other words, $\mathcal{S}_j>1$ only when the variance of one of the quadratures dips below the $\tfrac{1}{2}$-quantum level that is due to vacuum fluctuations in the ground state. We find a maximal $\mathcal{S}_2\approx1.2=0.7$\,dB ($\langle\hat{p}_2^2\rangle\approx0.43$) just before the instability threshold, which compares favourably to the amount of squeezing obtained in Ref.~\cite{Nunnenkamp2010} for pure quadratic coupling.\\
We conclude this Section by noting that the larger quadratic coupling strengths necessary for generating and observing squeezing can be obtained by making use of avoided crossings between pairs of cavity modes~\cite{Sankey2010}.

\section{Conclusions}
We have explored a system consisting of an optomechanical cavity with an additional intra-cavity particle whose motion is coupled quadratically to the cavity field. A direct coupling between the two moving elements arises naturally in this system. This coupling can be turned off and on at will by choosing which port the cavity is driven from, and therefore provides complete control over whether the interaction with the particle is purely quadratic, or whether it has some linear character. As examples of the potential of our scheme, we have discussed the possibility of cooling the motion of the sphere, despite its nominal quadratic coupling, as well as squeezing of this same motion. Our work opens up interesting perspectives for quantum state-transfer between two oscillators and the distribution of multipartite entanglement. Moreover, the system that we have addressed leaves room for asking interesting questions on the nonlocal nature of the three-mode state achieved via the coupling mechanisms addressed here, a goal that can be pursued, for instance, through the formal apparatus recently put forward in Ref.~\cite{Lee2012b}. 

\section*{Acknowledgements}
AX thanks the University of Malta for hospitality during completion of this work and the Royal Commission for the Exhibition of 1851 for financial support. MP is supported by the UK EPSRC through a Career Acceleration Fellowship and the ``New Directions for EPSRC Research Leaders'' initiative (EP/G004759/1). Parts of the calculations were carried out using computational facilities funded by the European Regional Development Fund, Project ERDF-080.

\appendix

\begin{figure}[t]
 \includegraphics[scale=0.5]{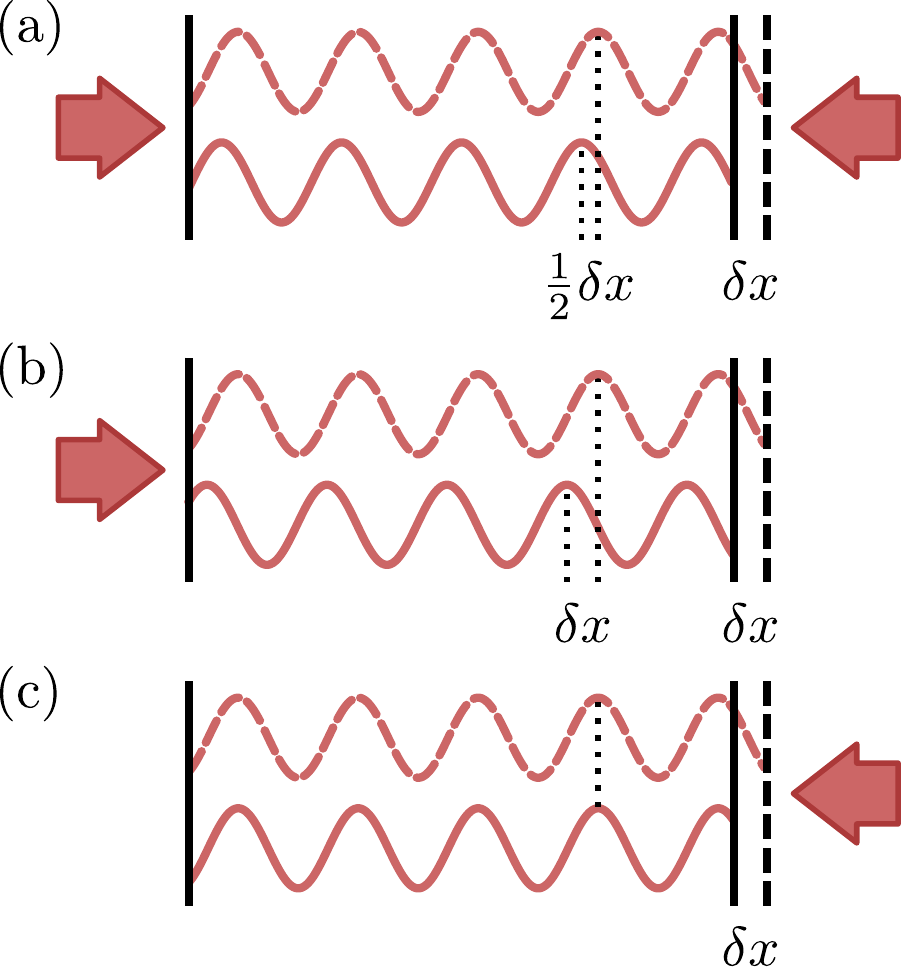}
 \caption{Different pumping geometries in one dimension. (a)~For symmetric pumping, a node or antinode always lies at the centre of the cavity. (b)~For pumping from the immobile mirror, the nodal structure shifts with the moving mirror. (c)~For pumping from the moving mirror, the nodal structure is fixed with respect to the immobile mirror.}
 \label{fig:Geometries}
\end{figure}
\section{Alternative pumping geometries}\label{app:Geometries}
The function of this Appendix is to elucidate the differences between pumping the cavity from the immobile mirror, from the moving mirror, or from both ends. These three possibilities, which are exhaustive in one-dimensional geometries, are illustrated in \fref{fig:Geometries}. The geometry used throughout this paper is shown in \fref{fig:Geometries}(b). The most `natural' case, where the cavity field is symmetric with respect to coordinate inversion, is depicted in \fref{fig:Geometries}(a). In both these situations, when the mirror on the right moves, the nodal structure shifts with respect to the laboratory frame, yielding a term in the Hamiltonian that couples the two oscillators. In the case of \fref{fig:Geometries}(c), however, the nodes of the field do not shift when the mirror moves, such that there is no longer any cross-coupling between the two oscillators. The case of a \emph{resonant} cavity pumped from one end is special:\ since all the input light is transmitted and none is reflected off the cavity, both mirrors effectively interact with three running waves of nonzero amplitude, two moving away from the pump, one towards it, such that the right mirror is a time-reversed copy of the left mirror. The entire system is therefore invariant under inversion of coordinates and time reversal.

In our notation, the part of the interaction Hamiltonian that involves the sphere can be represented as
\begin{equation*}
\hat{H}\propto\begin{cases}
\bigl(\tfrac{1}{2}\hat{x}_1-\hat{x}_2\bigr)^2\hat{a}^\dagger\hat{a} & \mathrm{(a)}\\
\bigl(\hat{x}_1-\hat{x}_2\bigr)^2\hat{a}^\dagger\hat{a} & \mathrm{(b)}\\
\hat{x}_2^2\hat{a}^\dagger\hat{a} & \mathrm{(c)}
\end{cases}
\end{equation*}
with reference to the respective cases in \fref{fig:Geometries}. The main text can be adapted for the remaining cases by setting $\chi\to\chi/2$ [case (a)] or $\chi\to0$ [case (c)].\par
This interesting result can easily be seen for case (b) (and analogously for the other two cases) by calculating the field inside the cavity, normalised to the input field, at a distance $z$ from the right mirror. That is
\begin{align*}
E(z)&=t\bigl[e^{ik(L-z)}+re^{ik(L+z)}+r^2e^{ik(3L-z)}+\dots\bigr]\nonumber\\
&=\frac{te^{ikL}}{1-r^2e^{2ikL}}\times(re^{ikz}+e^{-ikz})\nonumber\\
&=:\mathcal{L}(L)\times\mathcal{E}(z)\,,
\end{align*}
where $L$ is the length of the cavity, $k$ the wavenumber of the field, $\mathcal{L}(L)$ accounts for the spectral profile of the resonance, and $\mathcal{E}(z)$ depends only on the distance from the right mirror. For simplicity, the two mirrors are assumed to be identical, with (real) reflectivity $r$ and transmissivity $t$, and any phase shifts upon reflection or transmission are absorbed in $L$; these simplifications are not crucial to deriving this result. In the good-cavity limit $r\to-1$, $\mathcal{E}(z)$ reduces to the expected sinusoidal profile, with (anti)nodes lying at a fixed distance \emph{from the right mirror}.

\section{Drift matrix}\label{app:Drift}
Here we provide the explicit form of the drift matrix $\mat{A}$ for our problem. In order to aid comparison between datasets with different $g_1$ and $g_2$, we introduce the effective detuning $\tilde{\Delta}=\Delta+g_1\bar{x}_1-g_2\bigl(\chi\bar{x}_1-\bar{x}_2\bigr)^2$. The resulting drift matrix reads
\begin{widetext}
\begin{multline}
\label{eq:DriftMatrix}
\mat{A}=\left[
\begin{matrix}
-\kappa_\mathrm{c} & -\tilde{\Delta} & -\bigl[g_1-2g_2\chi\bigl(\chi\bar{x}_1-\bar{x}_2\bigr)\bigr]\bar{p} \\
\tilde{\Delta} & -\kappa_\mathrm{c} & \bigl[g_1-2g_2\chi\bigl(\chi\bar{x}_1-\bar{x}_2\bigr)\bigr]\bar{x} \\
0 & 0 & 0 \\
\bigl[g_1-2g_2\chi\bigl(\chi\bar{x}_1-\bar{x}_2\bigr)\bigr]\bar{x} & \bigl[g_1-2g_2\chi\bigl(\chi\bar{x}_1-\bar{x}_2\bigr)\bigr]\bar{p} & -\omega_1 - g_2\chi^2\bigl(\bar{x}^2+\bar{p}^2\bigr) \\
0 & 0 & 0 \\
2g_2\bigl(\chi\bar{x}_1-\bar{x}_2\bigr)\bar{x} & 2g_2\bigl(\chi\bar{x}_1-\bar{x}_2\bigr)\bar{p} & g_2\chi\bigl(\bar{x}^2+\bar{p}^2\bigr)
\end{matrix}\right.\qquad\cdots\\
\cdots\qquad\left.\begin{matrix}
0 & -2g_2\bigl(\chi\bar{x}_1-\bar{x}_2\bigr)\bar{p} & 0 \\
0 & 2g_2\bigl(\chi\bar{x}_1-\bar{x}_2\bigr)\bar{x} & 0 \\
\omega_1 & 0 & 0 \\
-2\gamma_1 & g_2\chi\bigl(\bar{x}^2+\bar{p}^2\bigr) & 0\\
0 & 0 & \omega_2 \\
0 & -\omega_2 - g_2\bigl(\bar{x}^2+\bar{p}^2\bigr) & -2\gamma_2
\end{matrix}
\right]\,.
\end{multline}
\end{widetext}
The phase of the input field can be chosen, as is commonly done, to set $\bar{p}=0$. One can recover the usual linear or quadratic optomechanics drift matrices by setting $g_2=0$ or $\chi=g_1=0$, respectively.

\end{document}